\documentclass[aps, showpacs,amsmath,amssymb,floatfix,twocolumn,prb,
10pt]{revtex4-1}
\usepackage{dcolumn}
\usepackage{amsmath,amssymb}
\usepackage{bm}
\usepackage{epsfig}

\begin{document}

\title{Pomeranchuk-Nematic instability in the presence of a weak magnetic field}

\author{Daniel  G.\  \surname{Barci} }
\affiliation{
Departamento de F\'{\i}sica Te\'{o}rica - Universidade do Estado do Rio de
Janeiro,
Rua S\~{a}o Francisco Xavier 524, 20550-013, Rio de Janeiro, RJ, Brazil.}

\author{Daniel \surname{Reyes} }
\affiliation{
Departamento de F\'{\i}sica Te\'{o}rica - Universidade do Estado do Rio de
Janeiro,
Rua S\~{a}o Francisco Xavier 524, 20550-013, Rio de Janeiro, RJ, Brazil.}

\date{\today}

\begin{abstract}
We analyze a two-dimensional Pomeranchuk-Nematic instability,
trigger by the Landau parameter $F_2<0$, in the presence of a small magnetic
field. Using Landau
Fermi liquid theory in the isotropic phase, we analyze the collective modes
near the quantum critical point $F_2=-1,\omega_c=0$ (where $\omega_c$ is the
cyclotron frequency). We focus on the effects of parity symmetry breaking on
the Fermi surface deformation. We show that, for  studying the critical regime,
the linear
response approximation of the Landau-Silin equation is not sufficient and it is
necessary to compute corrections at least of order $\omega_c^2$.
Identifying the slowest oscillation mode in the disordered phase, we compute
the phase diagram for the isotropic/nematic phase transition in terms of $F_2$
and $\omega_c$.
\end{abstract}




\maketitle

\section{Introduction}
The isotropic-nematic quantum phase transition was proposed as a possible
mechanism to explain the anisotropic behavior  of
several strongly correlated systems. Some interesting examples are  quantum Hall
liquids, high $T_c$ superconductors and  heavy fermions systems.
An interesting review can be found in reference~\onlinecite{EduardoReview}.

This transition can be understood as an instability of a Fermi surface under the
influence of a strongly  attractive two-body potential in the forward
scattering channel, with d-wave symmetry
(or equivalently, with angular momentum $\ell=2$). From the point of view of
Landau Fermi liquid theory, it is triggered by a Pomeranchuk instability
produced by a large negative value of the Landau parameter $F_2$ in the charged
sector.
As a consequence of the transition, the Fermi surface is deformed,
getting an  ellipsoidal component.
The
Goldstone modes, related with rotational symmetry breaking,   are dissipative
over-damped excitations,  characterized by dynamical exponent $z=3$.  The order
parameter theory was developed using different techniques: mean field
theory\cite{Ogan},
multidimensional bosonization\cite{BarciOxman, Lawler} and  Landau Fermi liquid
theory\cite{CastroNeto1}. While the
collective bosonic excitations are reasonably well understood, the fate of the
fermionic spectrum is still under debate\cite{Lawler, Sachdev, Kopietz1}.

From an experimental point of view, the study of Fermi surface deformations can
be performed by means of  at least two independent techniques: ARPES\cite{ARPES}
and the
observation of quantum oscillations\cite{QO},  like for instance, the de
Haas-van Alphen effect. The use of the latter resides in the ability to
reconstruct Fermi
surface shapes from the information contained in quantum oscillations of
different observables, when an externally applied magnetic field is varied.

The application of a strong magnetic field  suppresses any Pomeranchuk
instability, since it opens a gap in the spectrum due to Landau level
quantization. However, for small magnetic fields,  the Landau levels form a
dense set near the Fermi energy and strong attractive interactions mix all
levels in a non-trivial way.
Experimentally, nematic instability   have been
observed in the bilayer ruthenate compound
Sr$_{3}$Ru$_{2}$O$_{7}$ at
finite magnetic field~\cite{compound}, which
suggests that a meta-magnetic quantum critical point can be reached by changing
the direction of the applied magnetic
field~\cite{Perry,Borzi,Grigera}.
Therefore, it is important to understand the
critical behavior when the quantum critical point is reached by lowering the
magnetic field.

\begin{figure}
\centering
\includegraphics[height=5.5 cm]{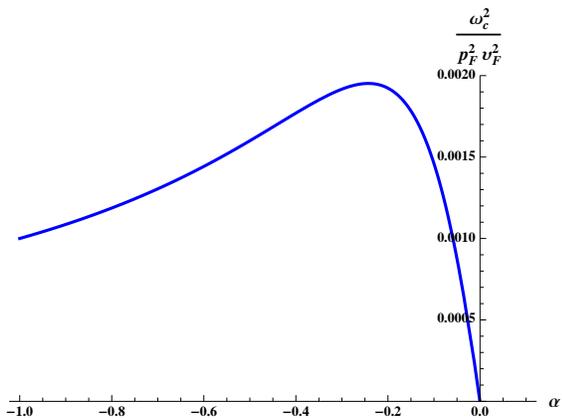}
\caption{Phase diagram for the Pomeranchuk instability $\ell=2$ in the presence
of a small magnetic field. The external
part of the transition curve represents an isotropic Fermi liquid, while the
inner part is an anisotropic  liquid phase. The adimensional control parameters
are $\alpha= 1 + F_2$ and   $\omega_c/\epsilon_F$, where $\omega_{c} = \frac{e
B}{m^{\star}}$ is the cyclotron frequency and $\epsilon_F=v_Fp_F$ is the Fermi
energy.  We have plotted equation (\ref{line}) by fixing the interaction range
$\kappa p_F=10$.}
 \label{pd}
 \end{figure}

With this motivation, we would like to present a study of a two-dimensional
Pomeranchuk-nematic instability in the presence of a small magnetic field,
applied perpendicular to the two-dimensional fermionic system. We have considered an
isotropic and homogeneous charged Fermi liquid  submitted to a small magnetic field,
$k_BT\ll\hbar \omega_c\ll \epsilon_F$, where $\omega_c=eB/m^{\star}$ is the cyclotron
frequency and $\epsilon_F$ is the Fermi energy of the system.
We have focused on a simplified model where only the attractive two-body
$d-$wave interaction is present.
Using a semi-classical approach, we have studied collective excitations of the
fermionic system using the Landau-Silin equation\cite{Landau,Silin}. Studying
the
oscillatory slowest mode, we can compute the transition line where the
isotropic phase gets unstable.
The main result is presented in figure (\ref{pd}) where we depict
the phase diagram for the nematic-Pomeranchuk instability. In this figure, the
horizontal axis is the usual Landau control parameter $\alpha=1+F_2$ while the
vertical axis is the adimensional magnetic field $(\omega_c/\epsilon_F)^2$.  We
observe a maximum value of the magnetic field above
which no Pomeranchuk instability is possible. Moreover,  we have observed a
reentrant behavior of the isotropic phase for greater
values of the interaction parameter.
We have also analyzed the behavior of collective modes near the
quantum critical point  ($F_2\to -1, \omega_c\to 0$).
Since the magnetic field breaks parity symmetry, the collective mode dynamics
mixes symmetric as well as antisymmetric modes. Then, the Fermi surface
deformation is not an ellipsoid but has a definite parity given by the
direction of the magnetic field and the momentum $\bf q$ of the periodic
perturbation.

The paper is structured as follows: in section \ref{SA} we briefly review the
Landau theory of
 charged Fermi liquids and the Landau-Silin equation to describe the collective
modes of a Fermi
 liquid submitted to an external magnetic field. In \S\ref{Model} we set our
model and deduce the phase diagram of figure (\ref{pd}). In \S\ref{CM} we
show the collective modes near the nematic quantum critical point.
Finally,  we discuss our results and we point out possible future developments
in
\S\ref{conclusions}.

\section{Semi-classical approximation}
\label{SA}

Following the standard Fermi liquid approach~\cite{Nozi} we start by writing down the energy functional
for a two-dimensional system of spin-less quasi-particles of effective mass
$m^{\star}$, in an
electromagnetic field defined by the  vector potential $\mathbf{A}=\mathbf{A}(\mathbf{r},t)$,
\begin{eqnarray}
E[n]&=&\sum_{\mathbf{p}}\frac{(\mathbf{p}+e\mathbf{A})^{2}}{2m^{\star}}n
(\mathbf{p},\mathbf{r})\nonumber\\
&+&\sum_{\mathbf{p},\mathbf{p}^{'}}\int d{\bf r} d{\bf r'}
f_{\mathbf{p}+e\mathbf{A};\mathbf{p'}+e\mathbf{A}}({\bf r}-{\bf r'})
\nonumber\\
&\times& n(\mathbf{p},\mathbf{r})n(\mathbf{p'},\mathbf{r})+
\mathcal{O}(n^{3}),
\end{eqnarray}
where $n(\mathbf{p},\mathbf{r})$ is the phase-space density at momentum $\mathbf{p}$ and position
$\mathbf{r}$.
$e$ is the quasi-particle charge and
$f_{\mathbf{p}+e\mathbf{A};\mathbf{p'}+e\mathbf{A}}({\bf r}-{\bf r'})$
is the Landau amplitude characterizing finite range two-particle interactions.
The Landau interaction function should depend on the electromagnetic vector
potential to guarantee gauge invariance.

In order  to compute a
semi-classical evolution equation, we define the effective single-particle
Hamiltonian
\begin{equation}
H_{\rm eff}(\mathbf{p},\mathbf{r})=\frac{\delta E[n]}{\delta
n(\mathbf{p},\mathbf{r})},
\label{Heff}
\end{equation}
which generates the following time evolution equation:
\begin{equation}
\frac{\partial n(\mathbf{p},\mathbf{r},t)}{\partial
t}=\left\{H_{eff},n(\mathbf{p},\mathbf{r},t)\right\}_{\rm PB}
+I_{\rm coll}[n(\mathbf{p},\mathbf{r},t)],
\label{evolution}
\end{equation}
where $\{\ldots\}_{\rm PB}$ are  Poisson brackets associated with the conjugate
variables ${\bf r}$  and ${\bf p}$ and the effects of quasi-particle scattering
are included in the collision integral $I_{\rm
coll}[n(\mathbf{p},\mathbf{r},t)]$.

By means  of the Hamilton's equations of motion
$d\mathbf{r}/d t=\nabla_\mathbf{p}
H_{\rm eff}(\mathbf{p},\mathbf{r},t)$ and $d\mathbf{p}/dt=-\nabla_\mathbf{r}
H_{\rm eff}(\mathbf{p},\mathbf{r},t)$,
and using equations (\ref{Heff}) and (\ref{evolution}),
it is  obtained the so-called Landau-Silin kinetic equation\cite{Silin,Nozi}
\begin{eqnarray}\label{kinetic}
&&\frac{\partial n(\mathbf{p},\mathbf{r},t)}{\partial t}+
\mathbf{v}(\mathbf{p},\mathbf{r},t)\cdot
\nabla_{\mathbf{r}}n(\mathbf{p},\mathbf{r},t)-
\left(  \mathbf{\mathcal{F}} (\mathbf{p},\mathbf{r},t)\right.  \nonumber\\
&&
\left.
+\sum_{\mathbf{p}^{'}}\int{d\mathbf{r'}}
f_{\mathbf{p}+e\mathbf{A};\mathbf{p'}+e\mathbf{A}}({\bf r}-{\bf r'})
\nabla_{\mathbf{r'}}n(\mathbf{p'},\mathbf{r'},t)\right)\cdot
\nonumber\\
&&\cdot\nabla_{\mathbf{p}
} n(\mathbf{p},\mathbf{r},t)=I_{\rm coll}[n(\mathbf{p},\mathbf{r},t)],
\end{eqnarray}
where
$\mathbf{\mathcal{F}}(\mathbf{p},\mathbf{r},t)=e[\mathbf{E}(\mathbf{r},
t)+\mathbf {v}(\mathbf{p},\mathbf{r},t)
\times \mathbf{B}(\mathbf{r},t)]$
is the Lorentz force and
$\mathbf{v}(\mathbf{p},\mathbf{r},t)=\nabla_{\mathbf{p}}H_{eff}$ is the
quasi-particle velocity, including interactions.
The Landau-Silin transport equation (\ref{kinetic}) resembles the conventional
classical Boltzmann equation. However, the effective Lorentz force
$\mathbf{\mathcal{F}}(\mathbf{k},\mathbf{r},t)$ depends
self-consistently on the quasi-particles distribution function $n(\mathbf{p},
\mathbf{r},t)$.  This evolution equation is the cornerstone of the present
work.

In this paper we want to  study the effect of an external magnetic field $B$,
applied perpendicular to the plane of the system.
We will assume that the cyclotron energy $\hbar
\omega_{c}=\hbar e  B/m^{\star}\ll\varepsilon_F$.
For simplicity through the paper we choose $\hbar\equiv1$.
In general, the scattering mechanisms described by the
collision integral can be studied applying the relaxation-time
$\tau$ approximation. We will consider that  the typical
collective mode frequencies are greater than the collision
quasi-particle frequency. Of course, this is not true at criticality.
However, to determine the position of the transition line,  it is enough to consider
$I_{\rm coll}\rightarrow 0$.
To set up the kinetic equation of the Fermi surface collectives modes,  let us
consider  a constant isotropic  equilibrium distribution $n_p^{0}$
and a small perturbation $\delta n$, such that
$
n(\mathbf{p},\mathbf{r},t)=n_{p}^{0}+\delta n(\mathbf{p},\mathbf{r},t).
$
In these conditions, the linear expansion of equation (\ref{kinetic}) in $\delta n$ provides
the transport equation
\begin{equation}\label{total}
\frac{\partial \delta n}{\partial
t}+\mathbf{v_p}^0\cdot\nabla_{\mathbf{r}}\delta
\bar n
-e[\mathbf{v_p}^0\times
\mathbf{B}]\cdot\nabla_{\mathbf{p}}\delta\bar{
n}=0\; ,
\end{equation}
where
\begin{eqnarray}\label{nbar}
\lefteqn{
\delta \bar{n}(\mathbf{p},\mathbf{r},t)=\delta n(\mathbf{p},\mathbf{r},t)-
\left(\frac{\partial n^{0}}{\partial
\varepsilon^{0}}\right) \times}
\nonumber \\
&\times&
\sum_{\mathbf{p'}}\int d
\mathbf{r'}f_{\mathbf{p}+e\mathbf{A};\mathbf{p'}+e\mathbf{A}}({\bf r}-{\bf
r'})\delta n(\mathbf{p'},\mathbf{r'},t)
\end{eqnarray}
is the deviation from local equilibrium.

It is important to point out that equation (\ref{total}) is linear in $\delta n_p$
however, is highly non-linear in the magnetic field since it enters the
definition of the Landau interaction (equation (\ref{nbar})).  Usually, to compute
collective plasma modes in charged Fermi liquids this last contribution is
neglected, ending with a true linear response theory\cite{Lee}. However, as we
will show,
this approximation is not consistent to  study Pomeranchuk instabilities.

At low temperatures $k_BT\ll \epsilon_{F}$ the electron dynamics is confined
to a small region around the Fermi surface. Then,  it is more
convenient to define
$
\delta n(\mathbf{p},\mathbf{r},t)=-(\partial
n_{\mathbf{p}}^{0}/\partial
\varepsilon_\mathbf{p})\nu_{\mathbf{p}}(\mathbf{r},t),
$
where $\nu_{\mathbf{p}}(\mathbf{r},t)$ measures local Fermi surface
deformation. Finally,
Fourier transforming in the space variable ${\bf r}$, the kinetic equation
(\ref{total}) becomes,
\begin{eqnarray}\label{total1}
&&\frac{\partial \nu_{\mathbf{p}}({\bf q},t)}{\partial
t}+\left(i\mathbf{v}_F^{0}\cdot{\bf q}
-e(\mathbf{v}_F^{0}\times\mathbf{B})\cdot\nabla_{\mathbf{p}}
\right)\nonumber\\
&&\times\left(\nu_{\mathbf{p}}({\bf q},t)+\delta\varepsilon_{\mathbf{p}}({\bf q}
, t)\right)=0,
\end{eqnarray}
where $\mathbf{v}_F^{0}$ is the Fermi velocity
and the expression
\begin{eqnarray} \label{epsilon}
\delta\varepsilon_{\mathbf{p}}({\bf q},t)&=&\frac{1}{V^2}\sum_{\mathbf{p'}}
\left(\frac{\partial n_{\mathbf{p'}}^0}{\partial
\varepsilon_\mathbf{p'}}\right)\int d{\bf r} d{\bf r'}\; e^{i  {\bf q}\cdot
{\bf r}}
\times \nonumber \\
&\times& f_{\mathbf{p}+e\mathbf{A},\mathbf{p'}+e\mathbf{A}}({\bf  r}-{\bf  r'})
 \nu_{\mathbf{p'}}({\bf r'},t)
\end{eqnarray}
describes the combined effect of interactions  and
magnetic field, being $V$ the space volume.

Equations (\ref{total1}) and (\ref{epsilon}) are the starting point of our
analysis.  They describe the dynamics of Fermi surface deformations, given an initial condition $\nu^{\rm in}({\bf q}, 0)$, representing
a small density fluctuation with wave-vector ${\bf q}$.
In the next section we will set up our model and will study the
Pomeranchuk instability in the nematic channel.

\section{The  Pomeranchuk Nematic Instability}
\label{Model}

For simplicity we consider a two-dimensional circular Fermi surface.  The
interaction Landau function  depends  essentially on the angle between two Fermi momenta and can be expanded in Landau parameters as
\begin{equation}
f_{\mathbf{p}+e\mathbf{A},\mathbf{p'}+e\mathbf{A}}({\bf  r}-{\bf
r'})\rightarrow f_{p_{F},p_{F}^{'}}(r)=\sum_\ell
f_\ell(r)e^{i\ell \varphi},
\end{equation}
where $\cos\varphi={\bf p}_F\cdot {\bf
p'}_F/p_F^2$.
Moreover, we can expand the deformation of the Fermi surface in
Fourier coefficients
\begin{equation}
\nu_{\mathbf{p}}(\mathbf{q},t)=\sum_\ell  \nu_\ell(q,t)
e^{i\ell\theta}
\end{equation}
where  $\cos\theta={\bf p}_F\cdot {\bf
q}/p_Fq$.

To study the Pomeranchuk-nematic instability, it is sufficient to
consider a simplified model defined by
$f_2({ r})\neq 0$, while $f_\ell(r) =0$ for all $\ell\neq 2$. The presence of
other interaction channels does not modify our results qualitatively, provided
they are all stable\cite{Lawler,CastroNeto1}.
We will consider a short-ranged but non-local interaction $f_2({ r})$,
whose Fourier transform is given by
\begin{equation}
\tilde f_2(q)=\frac{f_2}{1+ |F_2| (\kappa q)^2},
\end{equation}
where $F_2=N(0)f_2$ is the usual adimensional Landau parameter with angular
momentum $\ell=2$ ($N(0)$ is the density of states at the Fermi surface), and
$\kappa$ defines an effective interaction range $\xi=\sqrt{|F_2|}\kappa$.
Our approach is valid provided  $p_F^{-1}\ll\xi\ll q^{-1}$ \textit{i.e.},   when the interaction range is much larger than the
inter-particle distance, however shorter than the typical scale of long-ranged
perturbations.

In the absence of a magnetic field, the collective dynamics of the Fermi
surface, given by equation (\ref{total1}) with ${\bf A}=0$, reduces to
\begin{equation}\label{linear}
\frac{\partial \nu_\ell(q,t)}{\partial t}+\frac{i v_F
q}{2}\left[\alpha_{\ell-1}\nu_{\ell-1}(q,t)+\alpha_{\ell+1}\nu_{\ell+1}(q,
t)\right]=0,
\end{equation}
where we have defined $\alpha_{\ell}=1+F_{\ell}$ and
$F_\ell$ are
adimensional Landau parameters.
In our model $\alpha_2\equiv\alpha=1+F_2$ and $\alpha_\ell=1$ for all $\ell\neq
2$.

We can gain more physical insight by
defining symmetric and antisymmetric variables,
\begin{equation}
\nu^\pm_{\ell}(q,t)=\frac{1}{2}\left[\nu_{
\ell}(q, t)\pm \nu_{-\ell}(q,t)\right]
\end{equation}
in terms of which, the Fermi surface deformations are parametrized as
\begin{equation}
\nu({\bf q},\theta,t)=\sum_{\ell=0}^\infty \nu^+_{\ell}(q,t)
\cos(\ell\theta)+
\sum_{\ell=1}^\infty \nu^-_{\ell}(q,t) \sin(\ell\theta).
\end{equation}
Eliminating in equation (\ref{linear}) odd components in favor of  even ones,
we obtain the coupled
oscillator equations\cite{CastroNeto1}
\begin{eqnarray}\label{oscB0}
\lefteqn{
\frac{\partial^{2} \nu^\pm_{\ell}(q,t)}{\partial t^{2}}+\left(\frac{v_F
q}{2}\right)^{2} \left[A_\ell\; \nu^\pm_{\ell}(q,t)+ \right.} \nonumber \\
&+&\left. C_{\ell-1}\;\nu^\pm_{\ell-2}(q,t)+C_{\ell+1}\;\nu^\pm_{\ell+2}
(q,t)\right]=0
\end{eqnarray}
with the adimensional coefficients,
\begin{eqnarray}
A_{\ell}&=&\alpha_\ell(\alpha_{\ell-1}+\alpha_{\ell+1})\nonumber\\
C_{\ell}&=&\alpha_{\ell+1}\sqrt{\alpha_{\ell}\alpha_{\ell-1}}.
\end{eqnarray}
It is clear from  equation (\ref{oscB0}) that the even an odd
components of $\ell$ are decoupled. The same happens with the symmetric and
antisymmetric components. The physical reason for that is parity invariance.
Hence, the $\nu_2^+$ mode is coupled with the even symmetric modes $\nu_0,
\nu_4^+,\nu_6^+\ldots$. Near $F_2=-1$, or $\alpha\sim 0$, the $\nu_2^+$ mode
oscillates with frequency
\begin{equation} \label{omega2free}
\omega_2=\sqrt{2 \alpha}\left(\frac{v_Fq}{2}\right),
\end{equation}
while all the other modes essentially oscillate with $\omega_\ell\sim
v_Fq/\sqrt{2}$. Then, near $\alpha=0$, $\omega_2\ll \omega_{\ell}$ with
$\ell\neq 2$ showing that, in time scales $\tau\gg (v_Fq)^{-1}$,  $\nu_2^+$ is a
very slow mode, while all other rapid modes can be averaged to zero.
Therefore, when $\alpha\to 0$, the Fermi surface has an essentially elliptic
form during long periods of time. This is the onset of the Pomeranchuk-Nematic
instability.

When a magnetic field is applied,  parity, as well as time reversal symmetry
are broken. Then, the symmetric and antisymmetric modes are no longer
decoupled. In linear response theory, we can ignore the contribution of the
magnetic field in equation (\ref{epsilon}), then,  equation (\ref{total1}) can be
simplify to,
\begin{equation}\label{linearomega}
\frac{\partial \nu_\ell}{\partial t}+\frac{i v_F
q}{2}\left[\alpha_{\ell-1}\nu_{\ell-1}+\alpha_{\ell+1}\nu_{\ell+1}\right]
+i \ell\alpha_\ell\omega_c\nu_\ell=0,
\end{equation}
where we have defined the cyclotron frequency $\omega_c= eB/m^{\star}$.
Thus, the linear response correction to equation (\ref{linear}) is
proportional to $\alpha_\ell(\omega_c/v_F q)$, where $\alpha_\ell=1+F_\ell$.
Since $\alpha_\ell\sim 1$ for stable
modes ($\ell\neq 2$), this equation is suitable to study collective modes of the Fermi liquid
in small magnetic fields. However, near the Pomeranchuk instability
($\alpha_2\equiv\alpha\sim 0$),
$(\omega_c/v_F q)^2$ is of the same order of $\alpha(\omega_c/v_F q)$ and
cannot be ignored.
To see this more clearly,
we can compute the oscillation frequency of  $\nu_2^+$,  using equation
(\ref{linearomega}), obtaining,
\begin{equation}
\omega_2\sim\sqrt{2\alpha} \left(\frac{v_Fq}{2}\right) \left\{ 1+4\alpha
\left(\frac{\omega_c}{v_Fq}\right)^2+ \ldots\right\},
\end{equation}
where the ellipsis ``$\ldots$'' means terms of order $O
(\alpha^2(\omega_c/v_Fq)^4)$.
Clearly, for small $\alpha\ll 1$, the frequency is approximately given by
equation (\ref{omega2free}) without changing the behavior of the quantum critical point $\alpha=0$.

Therefore, to study the transition line $\omega_c(\alpha)$,  we need to
consider
quadratic corrections in the magnetic field.  To do this, we expand the Landau
function $f_2$ in equation (\ref{epsilon}),  keeping linear terms in the vector
potential ${\bf A}$,
\begin{eqnarray}
\lefteqn{
 f_{\mathbf{p}+e\mathbf{A},\mathbf{p'}+e\mathbf{A}}({\bf r}-{\bf r'})=
 f_2({\bf r}-{\bf r'}) \frac{({\bf p}\cdot {\bf p'})^2}{p_F^4} +}
\nonumber \\
&+&2e \frac{f_2({\bf r}-{\bf r'})}{p_F^4}({\bf p}\cdot {\bf p'}) \left[{\bf
A}({\bf r})\cdot ({\bf p}+{\bf p'}) \right]\;.
\end{eqnarray}
With this expression, equation (\ref{epsilon}) reduces to
\begin{equation}
\delta\varepsilon_{\mathbf{p}}({\bf q},t)=
\delta\varepsilon^0_{\mathbf{p}}({\bf q},t)+
\delta\varepsilon^A_{\mathbf{p}}({\bf q},t),
\end{equation}
where the first term has no contribution from the magnetic field, and is given
by
\begin{equation}
\delta\varepsilon^0_{\mathbf{p}}({\bf q},t)=-i
|f_2|
\sum_{\mathbf{p'}} \left(\frac{\partial n_{\mathbf{p'}}^0}{\partial
\varepsilon_\mathbf{p'}}\right)
\left(\frac{{\bf  p}\cdot {\bf p'}}{p_F^2}\right)^2\;
\nu_{\bf p'}({\bf q},t),
\end{equation}
while the second term is linear in $\omega_c$,
\begin{eqnarray}
\label{epsilonA}
\lefteqn{
\delta\varepsilon^A_{\mathbf{p}}({\bf q},t)=-2i
\left(\frac{\omega_c}{v_Fp_F}\right)
\frac{(\kappa p_F)^2 F_2^2}{p_F^4 N(0)} \; \times }\nonumber \\
&\times&\sum_{\mathbf{p'}} \left(\frac{\partial n_{\mathbf{p'}}^0}{\partial
\varepsilon_\mathbf{p'}}\right)
\left({\bf  p}\cdot {\bf p'}\right)\left[\left({\bf p} + {\bf p'}\right)\times
{\bf q} \right]\;
\nu_{\bf p'}({\bf q},t),
\end{eqnarray}
where we have chosen the symmetric gauge ${\bf A}=(1/2) {\bf r}\times {\bf B}$.
This term depends on the interaction range $(\kappa
p_F)^2$, then, for ultra-local interactions ($\kappa=0$), it makes no
contribution. On the other hand, the vectorial structure of  equation
(\ref{epsilonA}) filters  only contributions to  the  modes $\nu_{\pm 1},
\nu_{\pm 2}$.
Therefore, Fourier transforming in ${\bf p}$ we find for these modes,
\begin{eqnarray}
\label{nu1}
&&\frac{\partial\nu_1}{\partial t}+
i\left(\frac{v_Fq}{2}\right)\left\{
\left[1-2(1-\alpha)^2
\left(\frac{\omega_c}{v_Fq }\right)^2
(\kappa q)^2\right]\nu_0 \right.+ \nonumber \\
&&+\left.\left[\alpha-2(1-\alpha)^2
\left(\frac{\omega_c}{v_Fq }\right)^2
(\kappa q)^2\right]\nu_2\right\}   +i \omega_c\nu_1=0
\end{eqnarray}
and
\begin{eqnarray}
\label{nu2}
&&\frac{\partial\nu_2}{\partial t}+
i\left(\frac{v_Fq}{2}\right)\left\{1-4(1-\alpha)^2
\left(\frac{\omega_c}{v_Fq }\right)^2
(\kappa q)^2\right\}\nu_1+ \nonumber \\
&&+i\left(\frac{v_Fq}{2}\right) \nu_3+2i\alpha \omega_c\nu_2=0.
\end{eqnarray}
The equations for the modes $\nu_{-1}$ and $\nu_{-2}$ are easily obtained from equations
(\ref{nu1}) and (\ref{nu2}) by changing $\ell\to -\ell$, and  $\omega_c\to
-\omega_c$. The dynamical equations for the rest of the modes are simply given
by equation (\ref{linearomega}).
Then, building symmetric and antisymmetric mode combinations, and deriving
the evolution equations to get a second order system, we get for $\nu_2^+$,
\begin{eqnarray}
\label{nu2oscillator}
&&
 \frac{\partial^2\nu_2^+}{\partial t^2}+ \Omega^2 \nu^+_2+
 \left(\frac{v_Fq}{2}\right)^2\left(\nu_0+\nu_4^+\right)+ \nonumber\\
 &&+\left(\frac{v_Fq}{2}\right)\omega_c \left(2\nu_1^-+3\nu_3^-\right)=0
\end{eqnarray}
with
\begin{equation}
\label{omeg2interaction}
 \Omega^2=2\alpha\left(\frac{v_Fq}{2}\right)^2\left[
1+8\alpha\left(\frac{\omega_c}{v_Fq}\right)^2\right ]
+\left(\frac{\kappa q}{2}\right)^2(1-\alpha)^2 \omega_c^2.
\end{equation}
As we have anticipated, the magnetic field mixes symmetric and antisymmetric
modes.
The first contribution to the frequency in equation
(\ref{omeg2interaction}) comes from the linear
response theory, while the last term, proportional to the interaction range
$\kappa q \ll 1$,  is the first ``correction'' coming from equation
(\ref{epsilonA}).

Near the transition line $\Omega\to 0$, $\nu_0$ and  $\nu_4^+$ are  very
rapid and stable modes, while the coupling with the antisymmetric modes are very
weak. Thus, they do not modify the transition qualitatively. In the next
section we will study these couplings in more detail.
Therefore, $\nu_2^+$ is unstable when $\Omega=0$, leading to the
condition
line,
\begin{equation}
 \left(\frac{\omega_c}{v_F
p_F}\right)^2=-\frac{1}{8}\left(\frac{q}{p_F}\right)^2\frac{\alpha}{
\alpha^2+(1-\alpha)^2
(\frac{\kappa q}{4})^2}\; .
\label{qline}
 \end{equation}
 
Near the quantum critical point, we can expand this expression in powers of
$\alpha$,
\begin{equation}
 \left(\frac{\omega_c}{v_F
p_F}\right)^2=-2 \left(\frac{1}{\kappa p_F}\right)^2\alpha\ + O(\alpha^2)\; , 
\label{lienaralpha}
 \end{equation}
obtaining a linear critical region governed by the interaction 
range $\kappa p_F$.
Corrections of order $\alpha^2$ depend on $q$. Thus, 
differently from the usual Pomeranchuk transition, small perturbations with
different values of $q$ will contribute to the instability at different values
of $\alpha$. On the other hand, the momentum perturbation is limited to the
range $r_c^{-1}<q<\kappa^{-1}$. It is simple to show that the extremal line,
necessary to built up the complete phase diagram is reached at $q=1/\kappa$.
Therefore, the transition line is given by
\begin{equation}
\left(\frac{\omega_c}{v_F
p_F}\right)^2=-2\left(\frac{1}{\kappa p_F}\right)^2\frac{\alpha}{
16 \alpha^2+(1-\alpha)^2}\; , 
\label{line}
 \end{equation}
where the only free parameter is the interaction range $\kappa p_F>1$.
We depict eq. (\ref{line}) in figure (\ref{pd}).
As expected, a magnetic field strongly reduces the phase space for Pomeranchuk
instabilities. For small values of the magnetic field, the quantum critical point is
shifted to grater attractive values of the interaction $\alpha<0$, or $F_2<-1$.
Indeed, we observe a maximum value of the magnetic field
\begin{equation}
\left(\frac{\omega_c}{v_F p_F}\right)_{max}^2\sim 0.2
\left(\frac{1}{\kappa p_F}\right)^2
\label{eq.maximun}
\end{equation}
reached at
$
\alpha_{max}\sim -1/4,
$
above which, no Pomeranchuk instability is possible. Moreover,
we observe a reentrant behavior of the disordered isotropic phase for greater
values of the attractive interaction.

It is important  to note a clear difference with the case of the
usual Pomeranchuk instability. At zero magnetic field,
below the critical point $\alpha=0$, the isotropic Fermi liquid is unstable
under non-homogeneous density perturbations characterized by a wave-vector $q$.
Indeed, any value of $q$, no matter how small, will produce the phase
transition.
However, in the presence of a magnetic field, there is another  length-scale
given by the cyclotron radius $r_c=v_F/\omega_c$. This scale introduces
an infrared cut-off for the relevant fluctuations that could trigger the phase
transition. In other words, in the region below the
transition line in figure (\ref{pd}), the isotropic Fermi liquid is unstable
under density
fluctuations in a typical length-scale $\kappa<q^{-1}<r_c$.
In practice, $\kappa$ is a microscopic length and $r_c$ is very large, therefore
the above restriction is not
severe.  

On the other hand,
for $q^{-1}>>r_c$ there is no possible Pomeranchuk transition. This result is
quite clear. In the regime $q^{-1}>>r_c$, the semi-classical approach is no
more valid. It is necessary to treat the complete quantum problem, where the
system is gapped due to Landau level quantization. This is the quantum Hall
regime in which there is no Pomeranchuk instability.
From an experimental point of view, fluctuations and in particular, the
wave-vector $q$ are very difficult to control. However, any random
inhomogeneous density fluctuation will contain components of $ r_c^{-1}<q<
\kappa ^{-1}$ that,  even with a very small  amplitude, will trigger the
anisotropic/isotropic phase transition. On the other hand, it is always possible
to imagine (at least from a theoretical point of view) that one could induce a
small density fluctuation by applying a
modulated test field  with a  definite wave-vector $q$.

\section{Collective modes near  the quantum critical point}
\label{CM}

We would like to analyze the behavior of the stable oscillation modes of the
Fermi surface near the quantum critical point $(\alpha=0, \omega_c=0)$.
We are interested in the regime, $\alpha\ll 1$, and
$\omega_c\ll v_Fq \ll v_F p_F$.

We will focus in the unstable model $\nu_2^+$. This mode is directly coupled
with $\nu_0$, $\nu_4^+$, $\nu_1^-$ and $\nu_3^-$,  through equation
(\ref{nu2oscillator}). The symmetric modes $\nu_0$ and $\nu_4^+$
are stable modes and oscillate very rapidly
near the quantum critical point. Therefore, if we are interested in time scales larger
than $(v_Fq)^{-1}$, we can
average them to zero. The antisymmetric modes $\nu_1^-$ and $\nu_3^-$ couple
with $\nu_2^+$ through a magnetic field
$\omega_c$, as a manifestation of parity symmetry breaking. Then,  dismissing
the symmetric couplings,
$\nu_0$, $\nu_4^+$, the remaining system $(\nu_2^+, \nu_1^-, \nu_3^-)$ is a
closed one.  Defining the column vector
$\nu=(\nu_2^+, \nu_1^-, \nu_3^-)$, the collective modes satisfy,
\begin{equation}\label{linearsystem}
 \frac{\partial^2 \nu(q, t)}{\partial t^2}+ M \cdot\nu(q,t)=0,
\end{equation}
where the matrix $M$ takes the following form near the quantum critical point:
\begin{widetext}
\begin{equation}
M=\left(
\begin{array}{ccc}
 2\left(\frac{v_Fq}{2}\right)^2\alpha+ \omega_c^2(\frac{\kappa q}{2})^2
      &
\left(\frac{v_Fq}{2}\right) \omega_c
&
3\left(\frac{v_Fq}{2}\right) \omega_c  \\
      &  &    \\
\left(\frac{2\omega_c}{v_Fq}\right)\left[\left(\frac{v_Fq}{2}\right)^2\alpha+
 (\frac{\omega_c\kappa q}{2})^2\right]
 &
 \left(\frac{v_Fq}{2}\right)^2\alpha+
 \omega_c^2 (1+(\frac{\kappa q}{2})^2)
 &
 \left(\frac{v_Fq}{2}\right)^2\alpha+
 (\frac{\omega_c\kappa q}{2})^2 \\
 & & \\
 5\alpha\omega_c\left(\frac{v_Fq}{2}\right)
 & \alpha\left(\frac{v_Fq}{2}\right)^2
 & \left(\frac{v_Fq}{2}\right)^2
\end{array}
      \right)
      \; .
\end{equation}
\end{widetext}

It is instructive to analyze two different paths when approaching the quantum critical point.
In the case of zero applied magnetic field ($\omega_c=0, \alpha\to 0$), the
antisymmetric modes completely decouple
from the symmetric ones, due to parity symmetry. Then, the $\nu_2^+$
frequency  coincides with
equation (\ref{omega2free}). However, when approaching the quantum critical point
lowering the magnetic field ($\alpha=0, \omega_c\to 0$),
the matrix $M_c=\lim_{\alpha\to 0} M$ takes the form
\begin{equation}
M_c=\left(
\begin{array}{ccc}
\omega_c^2(\frac{\kappa q}{2})^2
      &
\left(\frac{v_Fq}{2}\right) \omega_c
&
3\left(\frac{v_Fq}{2}\right) \omega_c  \\
      &  &    \\
\left(\frac{2\omega_c}{v_Fq}\right)
 (\frac{\omega_c\kappa q}{2})^2
 &
 \omega_c^2 (1+(\frac{\kappa q}{2})^2)
 &
 (\frac{\omega_c\kappa q}{2})^2 \\
 & & \\
0
 &  0
 & \left(\frac{v_Fq}{2}\right)^2
\end{array}
      \right)\; .
\end{equation}

In order to find the normal modes, we diagonalize $M_c$  obtaining
the  eigen-values
\begin{equation}\label{eigenvalues}
\lambda_1=\left(\frac{\kappa q}{2}\right)^4 \omega_c^2,~~~~
\lambda_2=\omega_c^2,~~~~\lambda_3=\left(\frac{v_F q}{2}\right)^2,
\end{equation}
with the corresponding eigen-vector matrix
\begin{equation} \label{eigenvectors}
A=\left(
\begin{array}{ccc}
1
      &
2(\frac{\omega_c}{v_Fq})(\frac{\kappa q}{2})^2
&
0  \\
      &  &    \\
-2(\frac{\omega_c}{v_Fq})(\frac{\kappa q}{2})^2
 &
 1
 &
 0 \\
 & & \\
0
 &  0
 & 1
\end{array}
      \right) \;  .
\end{equation}
Thus, it is clear from equations (\ref{eigenvectors}) that the mode $\nu_3^-$
decouples for $\alpha=0$ and it is a rapid mode, oscillating with frequency
$v_Fq/2$. On the other hand, the modes $\nu_2^+$ and $\nu_1^-$ are slow modes,
coupled by the small quantities $(\omega_c/v_F q)\ll 1$ and $(\kappa q/2)^2\ll
1$. The former is related with the cyclotron frequency that should be smaller
than the frequency of a typical perturbation $(v_Fq)^{-1}$,  while the
latter is related with the interaction range that should be much smaller than
the typical length of the Fermi surface perturbation $q^{-1}$.

Therefore, very near the Pomeranchuk instability ($\alpha=0, \omega_c\ll
v_Fq$), the Fermi surface fluctuates following the equation
\begin{eqnarray}
\delta k_F&=& \nu_2^i \left\{ \cos(2\theta) + 2\left(\frac{\omega_c}{v_Fq}\right)\left(\frac{\kappa q}{2}\right)^2
\sin\theta \right\}\times \nonumber \\
&\times&\cos\left[\left(\frac{\kappa q}{2}\right)^2 \omega_c t+\varphi_1\right]
+ \nonumber \\
&+& \nu_{-1}^i \left\{ \sin(\theta) - 2\left(\frac{\omega_c}{v_Fq}\right)\left(\frac{\kappa q}{2}\right)^2
\cos(2\theta) \right\}\times \nonumber \\
&\times&\cos\left[\omega_c t+\varphi_2\right] + \nonumber \\
&+&\nu_{-3}^i \sin(3\theta) \cos\left[ (\frac{v_F q}{2}) t+\varphi_3\right],
\end{eqnarray}
where $\nu_2^i, \nu_{-1}^i, \nu_{-3}^i$ and $\varphi_1,\varphi_2, \varphi_3$ are
the initial amplitudes and phases, respectively.

We see that there are two slow modes that oscillate with frequencies
proportional to $\omega_c$. The slowest mode  ($\lambda_1$ in equation
(\ref{eigenvalues})) is related with  $\nu_2^+$, and it is responsible for
the Pomeranchuk instability when $\omega_c\to 0$. On the other hand, the mode
associated with the eigen-value $\lambda_2$ is related with the anti-symmetric
mode $\nu_1^-$. However, this mode is not unstable at the quantum critical point since,
when $\omega_c\to 0$, not only its frequency goes to zero, but
also its velocity $\partial \nu_1^-/\partial t\to 0$, implying a constant mode
at the quantum critical point, decoupled from any other symmetric mode.

\begin{figure}
\centering
\includegraphics[height=6 cm]{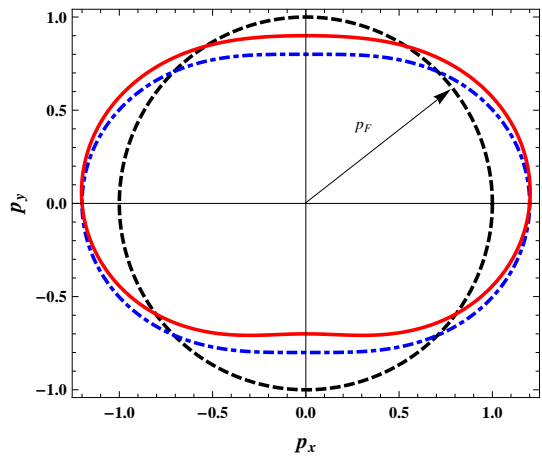}
\caption{ Snapshot of the Fermi surface deformation,  near the quantum critical
point
$\alpha=0, \omega_c=0$. The dash line is the actual Fermi surface, where we
have normalized $p_F=1$. The dash-dotted line is the elliptic deformation
without
magnetic field $\omega_c=0$. The continuous line is the deformation in the
presence of a small magnetic field. The vector ${\bf q}$ is aligned with the
$p_x$-axis, then  the parity symmetry breaking in the  $p_y$-axis is evident.}
 \label{SF}
 \end{figure}

In figure (\ref{SF}) we show a snapshot of the Fermi surface near the
Pomeranchuk instability, where we have chosen the initial conditions
$\nu_2^i=0.2$ and $\nu_{-1}^i=\nu_{-3}^i=0$. The circular dash line is the
actual isotropic Fermi surface. The dash-dotted line shows an ellipse, which
indicates the usual deformation with nematic symmetry in the absence of magnetic field,
while the continuous line is the deformation of the Fermi surface in the presence of a magnetic
field.
As expected, we observe a parity breaking in the axis ${\bf q}\times {\bf B}$ (
in this case in $p_y$ since we have chosen ${\bf q}$ pointing in the $p_x$
direction).  This is a tiny effect  proportional to
$(\omega_c/v_Fq)(\kappa q/2)^2\ll 1$. In
the figure we have artificially amplified this parameter,  in order to make
the effect of the magnetic field  visible.

We have re-done all the calculations of this section considering also the
couplings with the symmetric modes $\nu_0$ and $\nu_4^+$. In this case, it is
not possible to analytically solve the  resulting $9\times 9$ linear system.
However, making a numerical analysis,  we did not find any relevant deviation
from the simplified calculation shown. This confirms in some way that the
stable rapid modes do not participate in the instability process,  very near the
quantum critical point.

\section{Conclusions}
\label{conclusions}

We have analyzed the behavior of a two-dimensional Fermi liquid submitted to
an external magnetic field, near a Pomeranchuk instability triggered by the
Landau parameter $F_2$ in the charged sector.
We have considered a simple model in which the only interaction is given by
the Landau parameter $F_2$. The presence of other interactions does not modify
the results qualitatively, provided they are all stable, \textit{i.e.}, distant
from any other Pomeranchuk instability.

We have studied the Fermi surface stability, approaching the critical region
from
the isotropic phase, where the Landau theory of Fermi liquids can be used
safely. We have studied collective modes using the semi-classical
Landau-Silin equation. Usually, this equation was studied in the linear
response approximation to analyze plasma modes in charged Fermi liquids.
However,
near a Pomeranchuk instability this approximation is not sufficient. The reason
is that the quantum critical point is controlled by two parameters, $\alpha=1+F_2$ and
$\omega_c/v_F p_F$. The leading order correction in the magnetic field is
proportional
to $\alpha(\omega_c/v_F p_F)$. Thus, near the quantum critical point ($\alpha=0,
\omega_c=0$),
corrections proportional to  $\alpha^2$ and $(\omega_c/v_F p_F)^2$ are of the
same order
and cannot be neglected. Therefore, we need to go to quadratic order in the
magnetic field to consistently treat the neighborhood of the quantum critical point.

There are essentially three scales in the theory. The shortest distance
scale given by the inverse of the Fermi
momentum $p_F^{-1}$, an interaction range scale $\kappa$ and the longest
distance scale given by the cyclotron radius $r_c=v_F/\omega_c$.
We have found that the isotropic Fermi system could be  unstable under inhomogeneous
density fluctuations of typical length scale $q^{-1}$,  provided the inequality
$p_F^{-1}<\kappa\ll q^{-1}\lesssim r_c$  is satisfied.

Identifying the slowest collective mode, it is possible to compute
the transition line given in figure (\ref{pd}). The transition is completely
governed by the interaction range $\kappa p_F$. 
We observe an upper limit value
for  the magnetic field $\omega_c/v_Fp_F \sim 1/\kappa p_F$ over which the
Pomeranchuk instability is completely suppressed.  For smaller values of the
magnetic field, we observe that the instability is shift to stronger values of
the attractive interaction. Moreover, it is observed a reentrant behavior of the
isotropic phase for even stronger attractive interactions.
Reentrant behavior has posed challenges to microscopic theoretical
physics in a variety of condensed matter
systems~\cite{Fertig,Simanek,Cladis, Tinh,Berker, Manheimer,
Coutinho,Reyes1}. This phenomenon is characterized by the reappearance of
a less ordered phase, following a more ordered one, as a control parameter
(for example, temperature, pressure, chemical doping, magnetic field) is varied.
It appears that the re-entrance phenomenon also occurs, as we report in this
paper,
in the phase diagram for the Pomeranchuk  instability $\ell=2$ in the presence
of a small magnetic field. Basically, the reentrant phenomenon can be produced
by the increasing of entropy due to disorder or due to the presence of
additional degrees of freedom. We will leave the study of this new phenomena for a further work.

We have also studied collective modes couplings near the critical region. We
have identified the $\nu_2^+$ mode as the mainly unstable mode when the quantum critical point is approached.
The main contribution to the Fermi surface
deformation has elliptic (nematic) symmetry. However, the magnetic field couples
this mode with the antisymmetric ones $\nu_1^-$ and $\nu_3^-$. The
anti-symmetric  $\nu_3^-$ is a rapid mode oscillating with frequency $v_Fq/2$
and it does not participate of the instability process. On the other hand,
$\nu_1^-$ is a slow mode, however quicker than $\nu_2^+$,  since it oscillates
with the cyclotron frequency $\omega_c$. Even though its frequency goes to zero
at the quantum critical point, it does not represent a real Pomeranchuk instability
since, on one hand its coupling with $\nu_2^+$ also goes to zero with the
magnetic field, and not only its frequency but also its velocity goes to zero
as $\omega_c\to 0$.  However,  it has an important effect on the Fermi surface
deformation of the unstable mode since its coupling is a direct consequence of
parity breaking, producing a contribution that breaks nematic symmetry
as shown in figure (\ref{SF}). In fact, near the quantum critical point the
slowest mode is invariant under the combined transformation
$\theta\to \theta+ \pi, \omega_c\to -\omega_c$.

In order to have a complete picture of the isotropic-nematic phase transition
under the influence of a magnetic field, it is necessary to study  the ordered
phase. To do that in the context of Landau theory of Fermi liquids, it is
necessary to go beyond the linear approximation in $\delta n_p$ and to study the
collision integral $I_{\rm coll}$ in the Landau-Silin equation
(\ref{kinetic}).
Conversely, it is possible to face this problem with other approaches
like, for instance, non-perturbative calculations on specific fermionic models.
We hope to report on
this issue in the near future.

\acknowledgments
The Brazilian agencies, {\em Funda\c c\~ao de Amparo \`a Pesquisa do Rio
de Janeiro}, FAPERJ and {\em Conselho Nacional de Desenvolvimento Cient\'\i
fico e Tecnol\'ogico}, CNPq are acknowledged  for partial financial support. One
of us (D.R.) would like to thank  the {\em Aux\'{\i}lio
Intala\c{c}\~{a}o}-INST fellowship from FAPERJ for financial support.

\end{document}